\documentclass[apj]{emulateapj}
\usepackage{graphicx, graphics}
\usepackage{natbib}
\usepackage{amsmath,amssymb}

\received{}
\revised{}
\accepted{}

\newcommand{\fuse}{{\sl FUSE}}

\newcommand{\kms}{$\rm km\ s^{-1}$}

\shorttitle{3C~273}
\shortauthors{Fang \& Jiang}

\begin{document}

\title{High Resolution X-ray Spectroscopy of the Local Hot Gas along the 3C~273 Sightline}

\author{Taotao Fang\altaffilmark{1}, Xiaochuan Jiang\altaffilmark{1}}
\altaffiltext{1}{Department of Astronomy and Institute of Theoretical Physics and Astrophysics, Xiamen University; fangt@xmu.edu.cn}

\begin{abstract}

X-ray observations of highly ionized metal absorption lines at $z=0$ provide critical information of the hot gas distribution in and around the Milky Way. We present a study of more than ten-year {\sl Chandra} and {\sl XMM}-Newton observations of 3C~273, one of the brightest extragalactic X-ray sources. Compared with previous work, We obtain much tighter constraints of the physical properties of the X-ray absorber. We also find a large, non-thermal velocity at $\sim 100 - 150 \rm\ km\ s^{-1}$ is the main reason for the higher line equivalent width when compared with other sightlines. Using joint analysis with X-ray emission and ultraviolet observations, we derive a size of 5 -- 15 kpc and a temperature of (1.5--1.8)$\times10^6$ K for the X-ray absorber. The 3C~273 sightline passes through a number of Galactic structures, including the radio Loop I, IV, the North Polar Spur, and the neighborhood of the newly discovered ``{\sl Fermi} bubbles". We argue that the X-ray absorber is  unlikely associated with the nearby radio Loop I and IV; however, the non-thermal velocity can be naturally explained as the result of the expansion of the ``{\sl Fermi} bubbles". Our data implies an shock-expansion velocity of $200 - 300 \rm\ km\ s^{-1}$. Our study indicates a likely complex environment for the production of the Galactic X-ray absorbers along different sightlines, and highlights the significance of probing galactic feedback with high resolution X-ray spectroscopy.

\end{abstract}


\section{Introduction}

Since the launch of the {\sl Chandra} and {\sl XMM}-Newton X-ray telescopes, the on-board high resolution X-ay spectrometers open a new window of studying the interstellar medium (ISM) in our Galaxy. Numerous absorption features, produced by metal species with ionization stages ranging form neutral to hydrogen-like, were detected in the X-ray spectra of bright background sources such as the active galactic nuclei (AGNs) and Galactic X-ray binaries (XRB) (see, e.g., \citealp{nicastro2002, fang2003, rasmussen2003, yao2005, williams2005, bregman2007, pinto2010, hagihara2010, pinto2012, gorczyca2013, liao2013}). Due to the distance constraint, most XRBs can only probe the ISM within the Galactic disk, while AGNs can help study the ISM that may distribute beyond the disk and extend into the Galactic halo. AGN study of the Milky Way ISM, however, is constrained by the brightness of background sources. The most prominent absorption line produced by highly ionized metals at $z=0$ is the \ion{O}{7} K$\alpha$ transition at 21.6 \AA.\ So far, only a handful of targets are bright enough to detect other absorption line features, especially the crucial \ion{O}{7} K$\beta$ transition at 18.63 \AA,\ and enable the diagnose of the X-ray absorbing gas using high resolution X-ray spectroscopy. 

Mkn~421, PKS~2155 and 3C~273 are the three brightest extragalactic X-ray sources that have been repeatedly observed with {\sl Chandra} and {\sl XMM}-Newton. Despite sampling very different directions of Mkn~421 and PKS~2155-304, the two sightlines display a very similar distribution of hot gas (see \citealp{williams2005,williams2007,gupta2012}). The X-ray observation of the 3C~273 sightline, on the other hand, showed a quite different picture. Previous studies indicated that the detected \ion{O}{7} K$\alpha$ line EW along the 3C~273 sightlinee is about twice higher (see, e.g., \citealp{fang2003, rasmussen2003}). By comparing with the Mkn~421 sightline, \citet{yao2007a} concluded that the X-ray emission/absorption along the 3C~273 sightline line is enhanced by a Galactic Central Soft X-ray Enhancement (GCSXE) component.

\begin{figure*}[t]
\center
\includegraphics[height=0.45\textheight,width=1\textwidth]{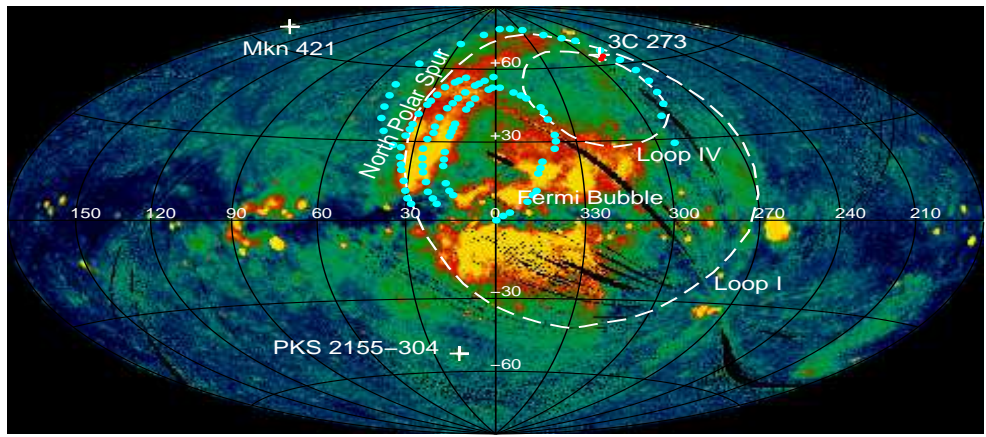}
\vskip-2cm
\caption{{\sl ROSAT} All-sky survey map (3/4 keV). Loop I and IV are the big and small dashed loops, respectively. North Polar Spur is the bright structure that starts at around $l=30^{\circ}$ near the Galactic plane and extends (nearly) perpendicular to above $b=60^{\circ}$. We label the direction of 3C~273, as well as two other AGNs, Mkn~421 and PKS~2155-304 with cross. The two red dots near the 3C~273 sightline are the {\sl XMM}-Newton pointings that we used for X-ray emission line analysis. The cyan circles are features identified in the {\sl Fermi}-LAT 1 -- 5 GeV data. These features include a giant bubble in the center, the northern arc (the two lines made up of cyan circles right next to the bubble) that extends up to $b\sim50^{\circ}$ and coincides with NPS, and a giant loop-shape structure that aligns with the radio Loop I and IV.}
\vskip0.4cm
\label{f1}
\end{figure*}

The sightline toward 3C~273 passes through part of the sky that is enriched with supernova (SN) activities (see the next section for details). Furthermore, since the early work of \citet{yao2007a}, two giant gamma-ray bubbles (``Fermi bubbles", see \citealp{su2010}) were discovered in the Galactic center, and the 3C~273 sightline passes through the edge of the northern bubble. Also, new data from both X-ray emission and absorption observations along this sightline were collected. Our goal of this paper is therefore to investigate the role of the Galactic structures along the 3C~273 sightline by making use of the latest data.

Our paper is organized as follows. In section \S2 we give a detailed description of the Galactic structures along the 3C~273 sightline. We perform the {\sl Chandra} data reduction and analysis in section \S3. In section \S4 we discuss the implication of our results by combining the results from the X-ray absorption observations with those from UV and X-ray emission observations. Last section is summary and discussion.

\section{The Sightline toward 3C 273}

The 3C~273 sightline passes through several complicated Galactic structures (see, e.g., \citealp{savage1993, sembach1997, sembach2001}). One of the most prominent structures in the radio continuum sky is the giant radio loops that covers a significant fraction of the sky. These radio loops are believed to be the superbubble structures created by a series of SN explosion. The 3C~273 sightline passes through the edges of two known radio loops, the Loop I and IV (see Figure~\ref{f1} for details). Loop I and IV are the result of a series of such explosions in the Sco-Cen OB association (\citealp{berkhuijsen1971,berkhuijsen1973}), and are about $\sim 130$ and 210 pc away, respectively. In Figure~\ref{f1} we plot the {\sl ROSAT} $3/4$ keV all-sky survey map in the Galactic coordinate \citep{snowden1995}. The origin and location of the prominent X-ray feature, the North Polar Spur (NPS), are still unclear. It has been suggested that the NPS could be the result of the Loop I superbubble colliding with the Local Hot Bubble (LHB) surrounding the Sun (see, e.g., \citealp{egger1995}), or a more distant phenomenon near the Galactic central region (see, e.g., \citealp{sofue2000, bland-hawthorn2003}).

Recently, two large gamma-ray bubbles (``{\sl Fermi} bubbles"), along with several large scale structures, were discovered in the {\sl Fermi}-LAT data (\citealp{su2010}, see our Figure~\ref{f1}). The giant northern bubble extends to $b\sim55^{\circ}$ ($\sim$ 10 kpc) above the Galactic plane, with a width of $\sim 40^{\circ}$ along the longitude direction. Next to the bubble is the northern arc (the two cyan dotted lines) that extends from $b\sim 0$ to nearly 60$^{\circ}$ and coincides with the prominent NPS feature in X-ray. {\sl Fermi} data also revealed a giant, loop-shape structure that aligns with the radio Loop I and IV. Recent studies indicated that the NPS may be related to the {\sl Fermi} bubbles, in particular, the northern arc (see, e.g., \citealp{su2010, guo2012, kataoka2013}). The 3C~273 sightline passes right through the outer loop-shape structure as well as part of the region that may associate with the northern arc and NPS. This sightline also passes through a region that is fairly close to the edge of the {\sl Fermi} bubbles.

\begin{deluxetable*}{lccccc}
\tablewidth{0pt}
\tabletypesize{\scriptsize}
\tiny
\tabletypesize{\footnotesize}
\tablecolumns{6}
\tablecaption{Observed $z\sim 0$ absorption lines\tablenotemark{a} \label{result}}
\tablehead{
\colhead{ID} &
\colhead{$\lambda_{\rm rest}$} &
\colhead{$v_{\rm obs}$} &
\colhead{$b$} &
\colhead{$EW$} &
\colhead{$\log N_i$} 
\\
\colhead{} &
\colhead{(\AA)} &
\colhead{(\kms)} &
\colhead{(\kms)} &
\colhead{(m\AA)} &
\colhead{$\rm cm^{-2}$} 
}
\startdata
\ion{O}{7} K$\alpha$ &21.602 &$119^{+28}_{-27}$&$129^{+27}_{-18}$ &$25.65\pm1.75$ &$16.36\pm0.10$ \\
\ion{O}{7} K$\beta$  &18.629 &-- &-- &$10.35\pm 1.43$ &-- \\
\ion{Ne}{9} K$\alpha$ &13.447 &$-43^{+58}_{-58}$ &$98^{+95}_{-54}$ &$12.05\pm 1.58$ &$15.92^{+0.28}_{-0.14}$ \\
\ion{Ne}{9} K$\beta$&11.547 &-- &-- &$1.44\pm 0.38$ &-- \\ 
\ion{O}{8} K$\alpha$ &18.969 &$170^{+100}_{-90}$& 133 &$7.07\pm 2.50$& $15.77^{+0.15}_{-0.20}$ 
\enddata
\tablenotetext{a}{All the errors are quoted at 1$\sigma$ level unless otherwise stated.}
\end{deluxetable*}

\section{{\sl Chandra} and {\sl XMM}-Newton Observations and Data Reduction}  

As a calibration source, 3C~273 ($z=0.158$, $l=289.95^{\circ}$, $b=64.36^{\circ}$) has been observed many times with {\sl Chandra} and {\sl XMM}-Newton. We focus on the observations performed with the high resolution gratings on-board of {\sl Chandra} and {\sl XMM}-Newton for a combination of high spectral resolution and reasonable sensitivities. In particularly, for {\sl Chandra} we analyzed the data obtained with the medium energy gratings (MEG) and low energy gratings (LEG), with the advanced CCD imaging spectrometer (ACIS) as the focal plane detector; for {\sl XMM}-Newton we focused on the Reflecting Grating Spectrometer unit 1 (RGS1); the other RGS unit, RGS2, has a failed CCD in the crucial \ion{O}{7} line region, so we did not use the data taken by RGS2.

Between 2000 and 2012 there are a total of 12 observations taken with MEG, and 6 observations with LEG, with total exposure time of 294 $ksec$ and 174 $ksec$, respectively. Each data set was analyzed following the standard procedure\footnote{\url{http://cxc.harvard.edu/ciao/threads/gspec.html}}, with version 4.5 of the {\sl Chandra} Interactive Analysis of Observations (CIAO) software. We co-added positive and negative first-order grating spectra to enhance the photon statistics. To simplify the spectral fitting, we also combined the 12 HETG observations and the 6 LETG observations to construct one MEG and one LEG spectrum. 

For {\sl XMM}-Newton between the same period, 20 observations of 3C~273 were performed with a total exposure of 769 $ksec$. Standard data reduction process was applied using the {\sl XMM}-Newton data analysis software SAS, version 12.0.1\footnote{See http://xmm.esa.int/sas/.}. We filtered periods that were impacted by high background events. Due to the large variation of the RGS1 response matrix between different observations, we cannot co-add all the data sets in a way that we adopted for the {\sl Chandra} data sets. All the data sets have to be fitted simultaneously to avoid systematic errors \citep{rasmussen2007}.

Using the X-ray spectral-fitting package $\it XSPEC$ version 12.7.1\footnote{See http://heasarc.nasa.gov/xanadu/xspec/}, the broadband continuum can be well-fitted by a power law plus the Galactic neutral hydrogen of $N_H=1.79 \times 10^{20}\rm\ cm^{-2}$~\citep{dickey1990} over $0.5 - 8.0$ keV. We clearly detected highly ionized \ion{O}{7} and \ion{Ne}{9} K-series (K$\alpha$, K$\beta$), and \ion{O}{8} K$\alpha$ transitions. For \ion{O}{7} K$\alpha$ and K$\beta$ transitions, we simultaneously fitted the LEG, MEG and RGS1 data with a Voigt profile-based line model. Due to a failed CCD in RGS1 around the  \ion{Ne}{9} line region and bad pixels presented in the RGS1 spectra around the \ion{O}{8} K$\alpha$ region, for these two ion species we used the {\sl Chandra} data only. The free parameters for this model are the line column density ($N_i$), the Doppler-$b$ parameter, and the redshift of the line center wavelength (see \citealp{buote2009} and \citealp{fang2010} for details). We also tied all these parameters together for the same ion but different transitions. In Table~1 we list all the measured and derived parameters. For the line $EW$, we estimated its error using Monte-carlo simulations \citep{fang2010}. We cannot constrain the Doppler-$b$ parameter of the \ion{O}{8} K$\alpha$ line, so we fixed the it at $b=129\rm\ km\ s^{-1}$, the best-fit value for the \ion{O}{7} line. 

\section{Analysis}

\subsection{Physical Properties Derived from X-ray Absorption Observations}

\begin{figure}[t]
\center
\includegraphics[height=0.25\textheight,width=0.45\textwidth]{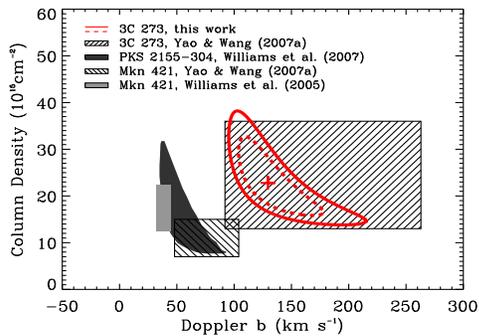}
\caption{Contour plot of \ion{O}{7} column density vs. the Doppler-$b$ parameter. The dashed and solid curves show the derived 1$\sigma$ and 90\% confidence contours for 3C~273, respectively, and the cross is the best-fit value. The hatched areas show the range of \ion{O}{7} column densities and the Doppler-$b$ parameters of 3C~273 and Mkn~421, measured in \citet{yao2007a}. Note that they adopted 90\% error range. We  also plot the contour for Mkn~421 (light grey square; \citealp{williams2005}) and PKS~2155-304 (dark-grey ellipse; \citealp{williams2007}) for comparison.}
\label{f2}
\end{figure}

The line $EW$ of an absorption line depends jointly on the ion column density and the Doppler-$b$ parameter. In Figure~\ref{f2} we show the column density vs Doppler-$b$ contour plot of the \ion{O}{7} K$\alpha$ line of 3C~273, as well as the comparison with other work. \citet{yao2007a} studied a subset of the {\sl Chandra} data that we analyzed in this paper. While our results are  largely consistent with theirs, compared with a total of $\sim 400\ ksec$ exposure in their work, better photon statistics has significantly improved the constraints on the key physical parameters. The measured 1$\sigma$ range of the Doppler-$b$ parameter is $31\ {\rm km\ s^{-1}} < b < 46\ \rm km\ s^{-1}$ for Mkn~421 \citep{williams2005} and $35\ {\rm km\ s^{-1}} < b < 94\ \rm km\ s^{-1}$ for PKS~2155-304 \citep{williams2007}. The column density in \citet{yao2007a} is somewhat lower than that of \citet{williams2007}, and we refer readers to \citet{yao2007} for a detailed discussion of the likely causes. We find that while for these three sight-lines the \ion{O}{7} column densities are similar on the order of $(1-2)\times10^{16}\rm\ cm^{-2}$, the Doppler-$b$ parameter of 3C~273, $129^{+27}_{-18}\rm\ km\ s^{-1}$, is significantly higher than those of PKS~2155-304 and Mkn~421, leading to a much higher line $EW$.

We now investigate the physical conditions of the X-ray absorber. A pure photoionization model is unlikely: Past modeling suggests that the observed column densities would require a typical intergalactic environment in which the path length of the absorber would be at $\sim$ Mpc level (see, e.g., \citealp{collins2005, williams2007}). We also consider the cases of collisional ionization. We adopted the ionization fractions calculated by \citet{gnat2007}, and that temperature of the X-ray absorber can be nicely constrained at a narrow range between $1.5\times 10^6$ K $ < T < 1.8\times10^6$ K, regardless the metal abundance and whether or not the gas is in collisional ionization equilibrium (CIE).

It is interesting to notice that assuming purely thermal broadening, this temperature would indicate a line width of $b_{th} = 0.129 \left(T/A\right)^{1/2}$, or $39\ {\rm km\ s^{-1}} < b_{th} < 43\ \rm km\ s^{-1}$. Here $A$ is the atomic weight. This line width is fully consistent with the measurements of the Mkn~421 and PKS~2155-304 absorbers, implying the X-ray absorbers along these two sightlines are most likely thermally broadened (see, e.g., \citealp{hagihara2010}). For 3C~273, the large Doppler-$b$ parameter suggests that non-thermal broadening plays a significant role. Since $b^2 = b^2_{th}+b^2_{nt}$, we derived a non-thermal velocity of $b_{nt} \sim 100 - 150\rm\ km\ s^{-1}$. 

\subsection{Joint Analysis with X-ray Emission and UV Observations}

Recently, \citet{henley2012} analyzed over thousands of {\sl XMM}-Newton observations and measured the \ion{O}{7} and \ion{O}{8} emission line intensities along these sightlines. In their sample, we select two pointings that are spatially close to the 3C~273 sightline, and use jointly the X-ray emission and absorption data to derive the physical properties of the X-ray absorber. The two pointings are located at $(l,b)=(289.172^{\circ}, 63.713^{\circ})$ ({\sl XMM} Obs ID: 0110990201), and $(l,b)=(292.777^{\circ}, 62.619^{\circ})$ ({\sl XMM} Obs ID: 0203170301), about 1$^{\circ}$ and 3$^{\circ}$ away from the 3C~273 sightline, respectively. Their measured \ion{O}{7} (\ion{O}{8}) line intensities are $14.17\pm1.27$ and $13.68\pm0.78$, respectively. Here L.U. is the line unit ($\rm photons\ cm^{-2}s^{-1}sr^{-1}$). Since $I_{\rm OVII} \propto Ln_e^2$, and $N_{\rm OVII} \propto Ln_e$, we find $n_e$ in the range of $\sim (1.2-1.8) \times10^{-3} \rm\ cm^{-3}$, and $L$ in the range $\sim (5 - 15)$ kpc. Here $n_e$ is the electron density, and $L$ is the size of the absorber. We also assume a solar abundance.

\citet{henley2012} measured the \ion{O}{8} emission line strength for these two pointings to be $2.46\pm2.31$ and $2.70\pm1.81$ L.U., respectively.  Adopting the above estimated values, we find the predicted \ion{O}{8} emission line intensity is $I_{\rm OVIII} \sim 1.9$ L.U., consistent with results from \citet{henley2012}. We also estimate an \ion{O}{8} column density of $\log N(\rm OVIII) (\rm cm^{-2})\sim 16.0$. This is slightly larger than what we observed, but not by much. We estimate an emission measure (EM) of $0.02\rm\ cm^{-6}\ pc$ along this direction. Again, this value is consistent with the EM derived in \citet{yao2007a}, in which they used {\sl ROSAT} data.

\begin{figure}[t]
\center
\includegraphics[height=0.25\textheight,width=0.45\textwidth]{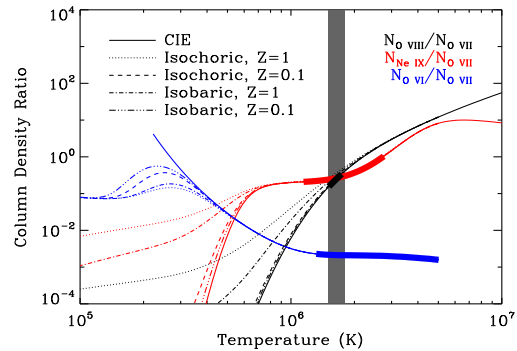}
\caption{Column density ratio between \ion{O}{8} and \ion{O}{7} (black), \ion{Ne}{9} and \ion{O}{7} (red), and \ion{O}{6} and \ion{O}{7} (blue), as a function of temperature. The solid line is for the collisional ionization equilibrium (CIE). The dotted and dashed lines represent the isochoric cooling for 1 and 0.1 solar abundances; and the dot-dashed, dot-dot-dot-dashed lines are for the isobaric cooling. The thick line segments indicate the constrains from observations, and the vertical shadowed area show the temperature region that fit all observations.}
\label{f3}
\end{figure}

\citet{gupta2013} noticed that while the absorption line systems are remarkably similar between Mkn~421 and PKS~2155-304 sightlines, their X-ray EMs are different by a factor of $\sim$ 1.5 ($\sim 0.0025\rm\ cm^{-6}\ pc$ for Mkn~421 and $\sim 0.0042\rm\ cm^{-6}\ pc$ for PKS 2155-304). They attributed this to the anisotropy of the circum-galactic medium (CGM), for which the Mkn~421 absorber has a much larger path length. Our estimated EM is much larger than both sightlines. Using their model the 3C~273 absorber would have a path length longer than the size of the Local Group. So it appears more likely the 3C~273 absorber is local around the Galactic disk, with a high density and a strong non-thermal motion.

\fuse\ observed 3C~273 for 43.3 $ksec$ on April 23, 2000 (see, e.g., \citealp{sembach2001, wakker2003, sembach2003}). The strong, $\lambda$ 1031.93 \AA\ transition was clearly detected with three distinguish components, with velocities at -160 -- 100, 105 -- 160, as well as a broad wing at 160 -- 260 $\rm km\ s^{-1}$. We analyzed the three components and found for the broad-wing component, the same collisional ionization can explain the column density ratio between \ion{O}{7} and \ion{O}{6} very well (the blue lines in Figure~\ref{f3}). This component has an \ion{O}{6} column density of $\log N(\rm O~VI) \sim 13.52$.

\subsection{Possible Association with the ``{\sl Fermi} bubbles"}

The derived size of the X-ray absorber along 3C~273 sightline, $\sim$ 5 -- 15 kpc, firmly puts the hot gas at the distant Galactic disk and possibly beyond, toward the direction of the Galactic central region. The X-ray absorber is unlikely to  be associated with either radio Loop I or IV, since both structures are within a few hundred pc; it is also unlikely to be linked to the giant loop-shape structure seen in the {\sl Fermi}-LAT data (see Figure~\ref{f1}). Also, the EM along this sightline is nearly an order-of-magnitude higher than those measured along the Mkn~421 and PKS~2155-304 sightline. While \citet{gupta2013} suggested the hot gas along these two sightlines is located at the distant Galactic halo with a path length of more than 100 kpc, it is unlikely the case for the 3C~273 sightline.  

This sightline passes through part of the region that is associated with the NPS and the northern arc discovered in the {\sl Fermi}-LAT data. Considering the proximity to the northern {\sl Fermi} bubble, it is interesting to discuss the impact of the {\sl Fermi} bubbles on the X-ray absorber along the 3C~273 sightline. It is unlikely the X-ay absorption/emission is produced by the same material located within the {\sl Fermi} bubbles: the temperature inside the bubble is hotter than $10^7$ K \citep{su2010}, and all the ion species we detected in the X-ray would be completely ionized at this temperature.

However, such X-ray absorber can be located in the region that is shock-impacted by the expansion of the {\sl Fermi} northern bubble. X-ray observations of the edge of {\sl Fermi} bubble suggested a possible link with the NPS in X-ray \citep{kataoka2013}. Assuming an adiabatic gas of $\gamma=5/3$, we estimated a shock Mach number of $\sim2$ and a shock expansion velocity of $v_s = 2b_{nt} \sim 200 - 300 \rm\ km\ s^{-1}$, following \citet{kataoka2013}. Such velocity would imply a bubble formation time of $(3-5) \times 10^7$ yr, assuming a bubble size of $\sim$ 10 kpc. The shock temperature for an adiabatic expansion shock can be estimated as $T_s = 1.34\times10^5 (v_s/100)^2 \sim 10^6$ K \citep{stocke2006} for $v_s = 300\rm\ km\ s^{-1}$, close to the temperature predicted by the UV and X-ray ion ratios in Figure~\ref{f3}.

\section{Summary and Discussion}

In this paper, we have analyzed the {\sl Chandra} and {\sl XMM}-Newton grating observations of 3C~273, one of the brightest extragalactic X-ray sources, with a focus on the X-ray absorption lines produced by highly ionized metal species at $z=0$. We summarize our finding below.

Using high resolution X-ray spectroscopy we measured the physical properties of the X-ray absorber along the 3C~273 sightline. Our measured line properties are largely consistent with previous work of \citet{yao2007a} based on a subset of the data used in this paper, while better photon statistics allows us to put much tighter constraints. The column density ratios between \ion{O}{7}, \ion{O}{8}, and \ion{Ne}{9} suggested a temperature of $(1.5 - 1.8) \times10^6$ K of the X-ray absorbing gas. This gas is either in collisional ionization equilibrium, or cooling at constant density or constant pressure. A joint analysis with the X-ray emission data suggests that the X-ray absorber likely has a density of $(1.2 - 1.8) \times 10^{-3}\rm\ cm^{-3}$, with a linear size of $\sim$ 5 to 15 kpc.

We compare the 3C 273 sightline with the sightlines of Mkn~421 and PKS~2155-304, which have similar or even better quality X-ray spectra. We find the line $EW$s of the 3C~273 are significantly higher than the other two sightlines. In particular, we find the large $EW$s are the result of higher Doppler-$b$ parameter, rather than large column density. While the Doppler-$b$ parameters of the X-ray absorbers detected in the Mkn~421 and PKS~2155-304 sightlines can be naturally explained by thermal broadening, a significant non-thermal component of $b_{nt} \sim 100 - 150\rm\ km\ s^{-1}$ is presented in the 3C~273 X-ray absorber. 

Such non-thermal velocity has been suggested before as an evidence of an outflow, possibly produced by stellar wind and supernova activities in the Galactic center and the bulge region \citep{yao2007a}. Large scale, bi-polar outflow was also proposed based on {\sl ROSAT} and other observations (see, \citealp{sofue2000, bland-hawthorn2003}). We suggest here an alternative that the non-thermal velocity we detected in the 3C~273 sightline can be naturally explained as the expansion of the newly discovered ``Fermi bubble", with a shock velocity of $\sim 200 - 300\rm\ km\ s^{-1}$. The derived expansion velocity is much less than predictions from recent theoretical modeling (see, \citealp{guo2012, yang2013}), as suggested by \citet{kataoka2013}.

\acknowledgments
We thank for helpful discussion with Charles Danforth, Fulai Guo, and Junfeng Wang. We also thank the referee for very helpful suggestions. TF was partially supported by the National Natural Science Foundation of China under grant No.~11243001 and No.~11273021, also by ``the Fundamental Research Funds for the Central Universities" No.~2013121008.


\end{document}